\documentclass{PoS}

\graphicspath{{figs/}}
\usepackage{url}
\usepackage{heppennames2}
\usepackage{placeins}
\providecommand{\mhpm}{m_{\PSHpm}}
\providecommand{\mhmodp}{\ensuremath{m_{\Ph}^{\rm mod+}}}
\newcommand{\Pt}{{\mathrm{t}}}

\title{Summary of recent progress in searches for additional Higgs bosons}

\ShortTitle{Summary of recent progress in searches for additional Higgs bosons}

\author{\speaker{Martin Flechl}\\
        Institute of High Energy Physics,\\
        Austrian Academy of Sciences\\ 
        E-mail: \email{martin.flechl@cern.ch}}

\abstract{
There has been a tremendous increase of sensitivity in searches for charged Higgs bosons and additional neutral Higgs bosons 
since the last CHARGED workshop in 2016. We will review recent experimental and theory developments as presented at CHARGED2018, 
and conclude with future prospects for the field.
}

\FullConference{
Prospects for Charged Higgs Discovery at Colliders - CHARGED2018\\
		25-28 September 2018\\
		Uppsala, Sweden}

\begin{document}

\section{Introduction}
Additional Higgs bosons $\PSHpm$ appear in many extensions of the standard model (SM),
in particular when adding additional doublets or triplets to its scalar sector. Typically, 
the focus is on 2-Higgs-doublet models (2HDMs) including the special case of the Higgs sector of the minimal supersymmetric extension of the standard model (MSSM). 
In the MSSM, the dominant production mode for a charged Higgs boson is in top quark decays if kinematically allowed, or in association with a top quark 
otherwise. Neutral Higgs bosons are dominantly produced via gluon fusion (for low values of $\tan \beta$, the ratio of the vacuum expectation values of 
the two Higgs doublets) and in association with bottom quarks (intermediate and high $\tan \beta$). However, non-standard production and decay modes may 
become dominant in other 2HDMs or for specific relations between the Higgs boson masses, allowing e.g. Higgs-to-Higgs decays. Other 
beyond-the-standard-model (BSM) Higgs sectors offer an even larger variety: for example, the next-to-MSSM (NMSSM) predicts two neutral Higgs bosons in 
addition to the MSSM which could be relatively light without violating existing bounds; and models with Higgs triplets predict among others doubly-charged Higgs bosons.

\section{Charged Higgs bosons}
\begin{figure}[htb]
\begin{center}
\includegraphics[width=.32\textwidth]{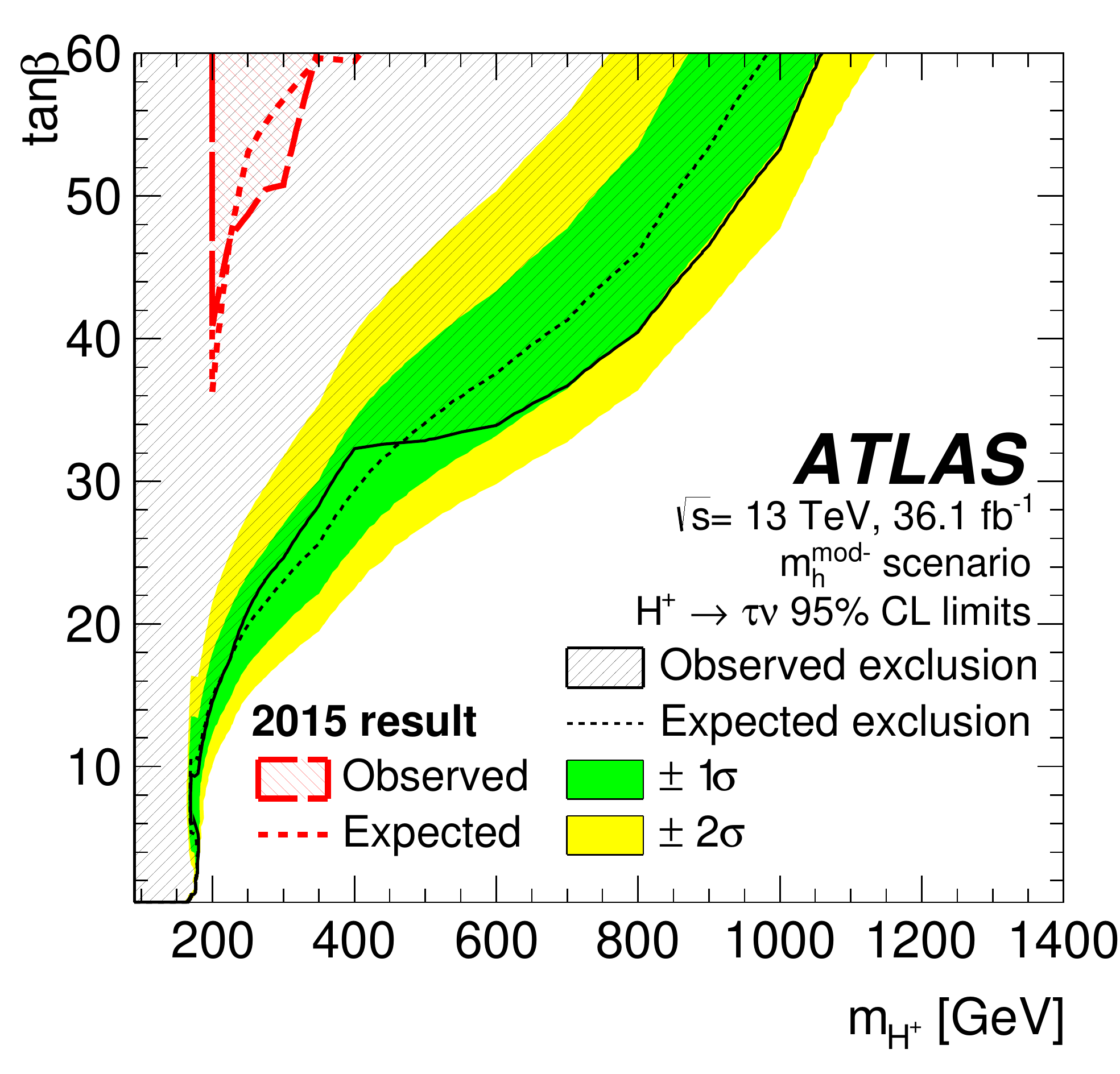}
\includegraphics[width=.32\textwidth]{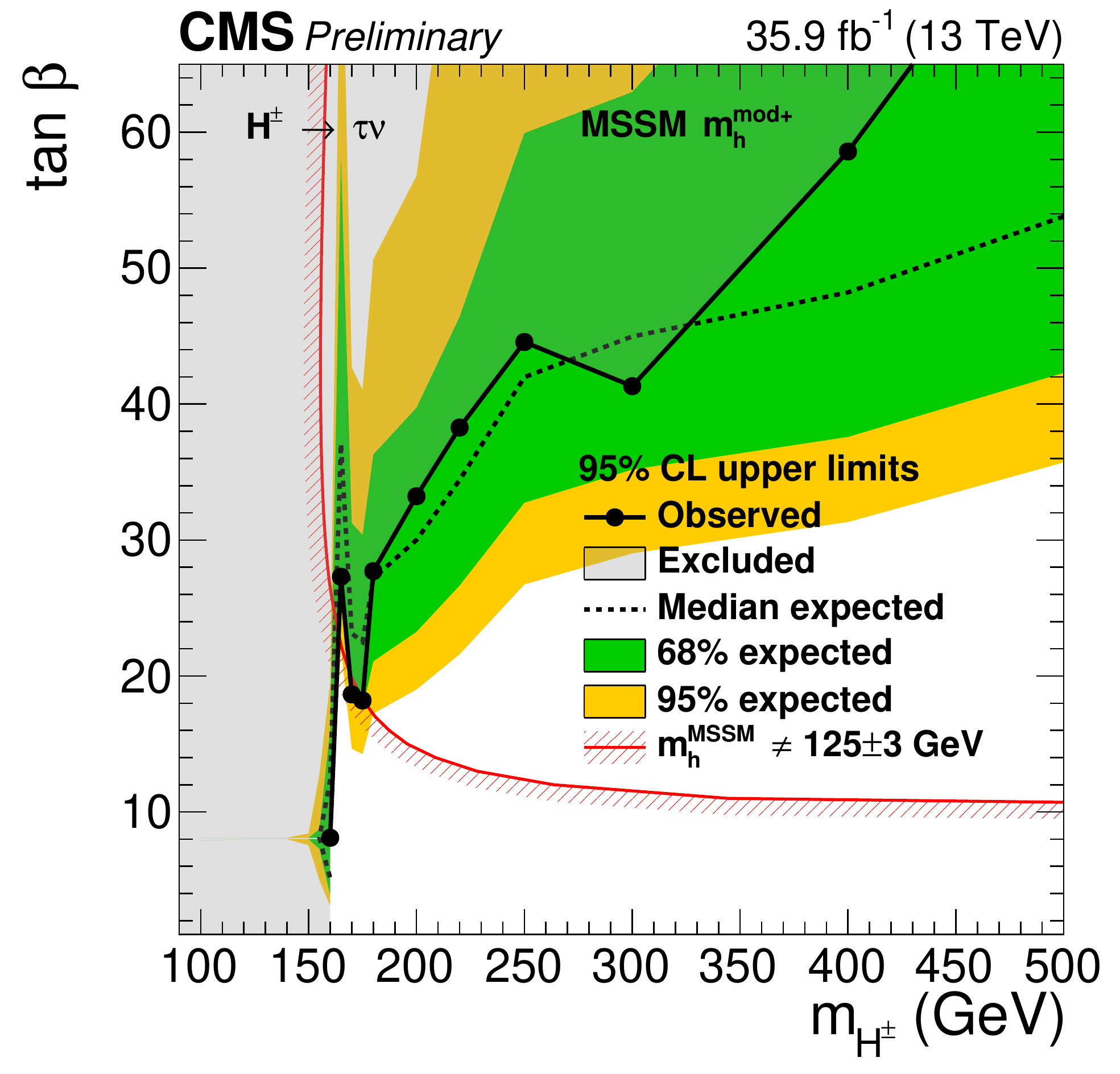}
\includegraphics[width=.32\textwidth]{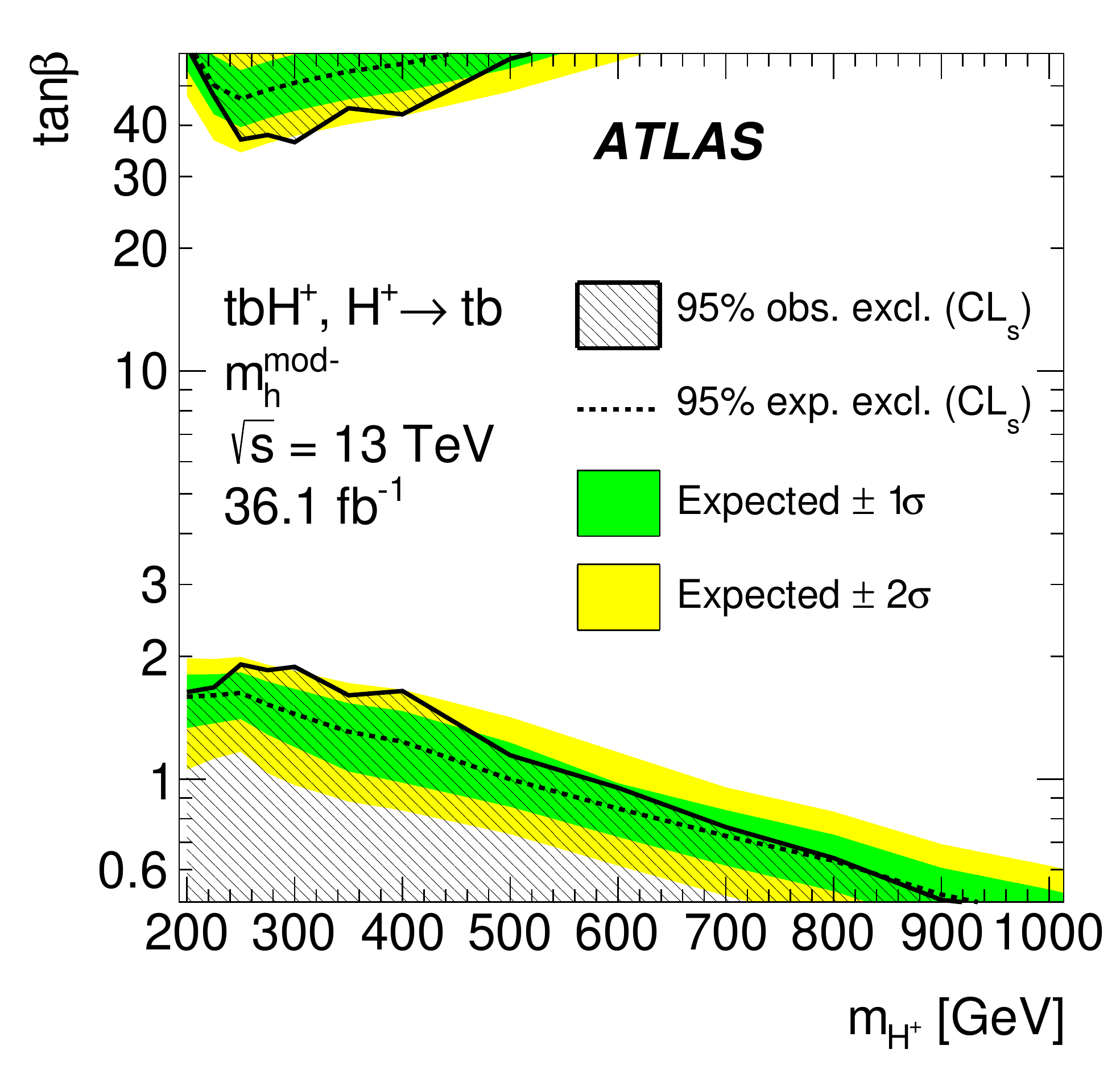}
\caption{
95\% CL limits for the MSSM interpretation of the search for $H^\pm \to \tau\nu$ by ATLAS~\cite{atlas_hp_tau} (left) and CMS~\cite{cms_hp_tau} (middle), and 
of $H^\pm \to tb$ by ATLAS~\cite{atlas_hp_tb} (right).
}\label{hp}
\end{center}
\end{figure}
Charged Higgs boson ($H^\pm$) searches are primarily motivated by the MSSM but the results are also given in a largely model-independent way, i.e. in terms of 
cross section limits. There is also an increasing number of non-MSSM searches, e.g. for charged Higgs boson production in vector boson fusion~\cite{cms_vbf_hp}. 
For masses below the top quark mass, a charged Higgs boson
would be dominantly produced in top quark decays. Therefore, the production cross section is proportional to the top quark pair production times
the branching ratio $\Pt \to \PH^+ \Pb$.
The dominant mode of $\PSHpm$ production for $\mhpm > m_\Pt$ in a 2HDM is via the process $\Pp\Pp\, \rightarrow\, \Pt\PSHpm + X$~\cite{flechl,berger,dittmaier,weydert,yr4}.
Recent progress on theory side has led to the first consistent NLO calculations also for the so-called intermediate-mass region~\cite{degrande}, i.e. 
the region where the contributions with and without intermediate on-shell top quarks are of similar size. This region is of particular interest because MSSM models 
where the heavy neutral scalar has a mass of 125 GeV and the light Higgs boson can act as mediator to dark matter typically have a charged Higgs boson mass in 
this region.

In the MSSM, the most sensitive $H^\pm$ decay mode is to a $\tau$ lepton and a neutrino~\cite{atlas_hp_tau,cms_hp_tau}, except for very low 
values of $\tan \beta$. If kinematically allowed, decays to $tb$ are typically similarly abundant, however, reconstructing 
such events is experimentally even more challenging~\cite{atlas_hp_tb}. An MSSM interpretation of the obtained limits is given in Fig.~\ref{hp}. Previous limits have 
been extended from several hundreds of GeV to one TeV (for the same value of $\tan \beta$) in the last two years, marking tremendous progress since 
the last workshop. These results also mark the first results for the intermediate-mass region.

Several other ways to search for charged Higgs bosons have been proposed to the experimental collaborations at this workshop. Examples are 
charged Higgs boson decays to $\PW\gamma$~\cite{logan}, $H^\pm$ production in $\PQc\PAQs$ fusion in a 3HDM~\cite{pasechnik}, or 
charged Higgs boson decays to neutralinos and charginos~\cite{bahl}.

\section{Doubly-charged Higgs bosons}
Doubly-charged Higgs bosons $H^{\pm\pm}$ are for example predicted by models with Higgs triplets such as left-right symmetric models or the Georgi-Machacek model, 
and the Zee-Babu model which only adds two singlets to the SM. Variants of these models can serve to explain the tiny observed neutrino masses or 
to restore parity symmetry in weak interactions at high energy. Models with triplets do not automatically lead to the observed value of $\rho \equiv \frac{M_W^2}{cos^2 \theta_W M_Z^2}$ close to unity, as models with only additional singlets and doublets do, and are thus more or less severely constrained.

\begin{figure}[htb]
\begin{center}
\includegraphics[height=3.2cm]{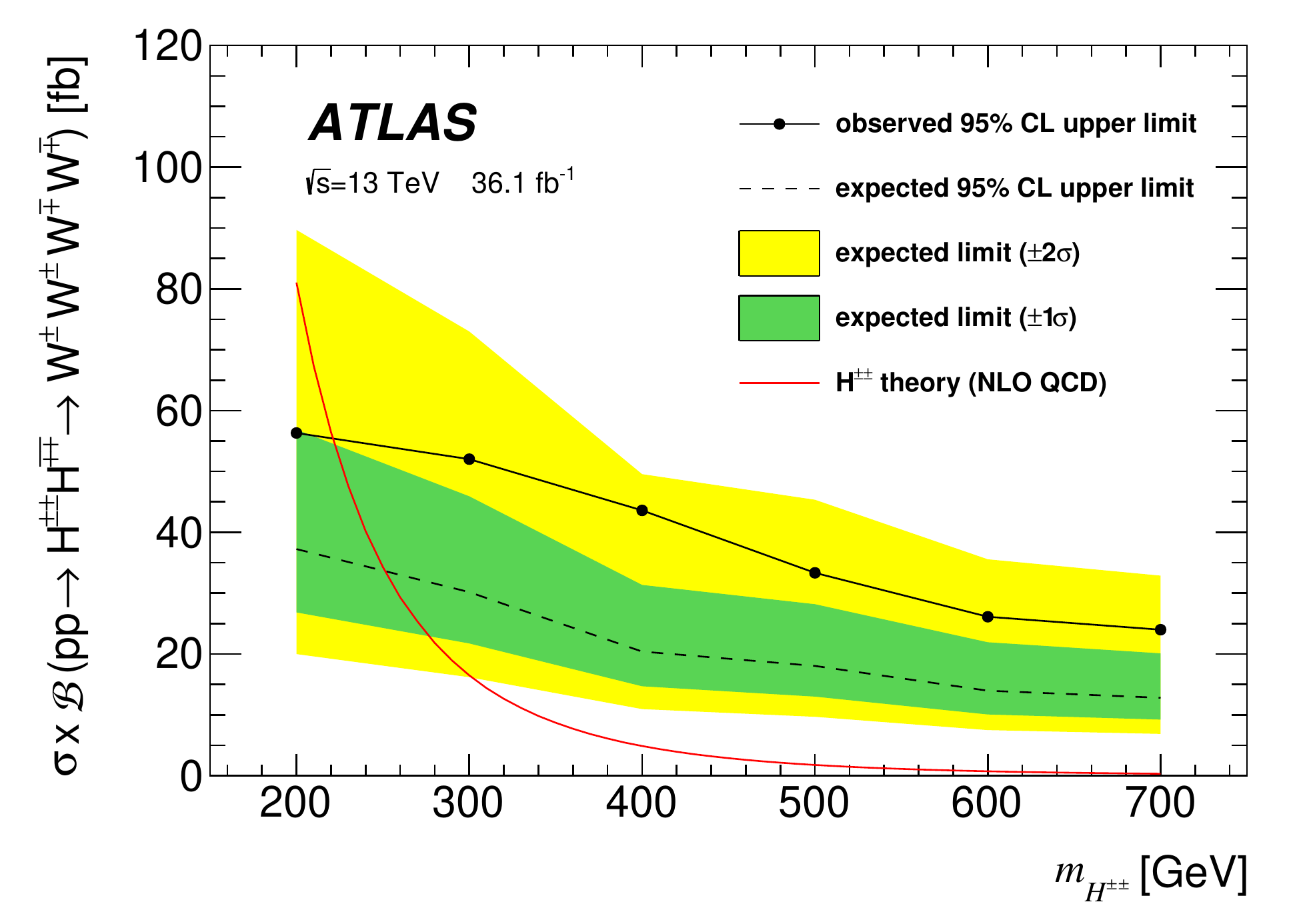}
\includegraphics[height=3.2cm]{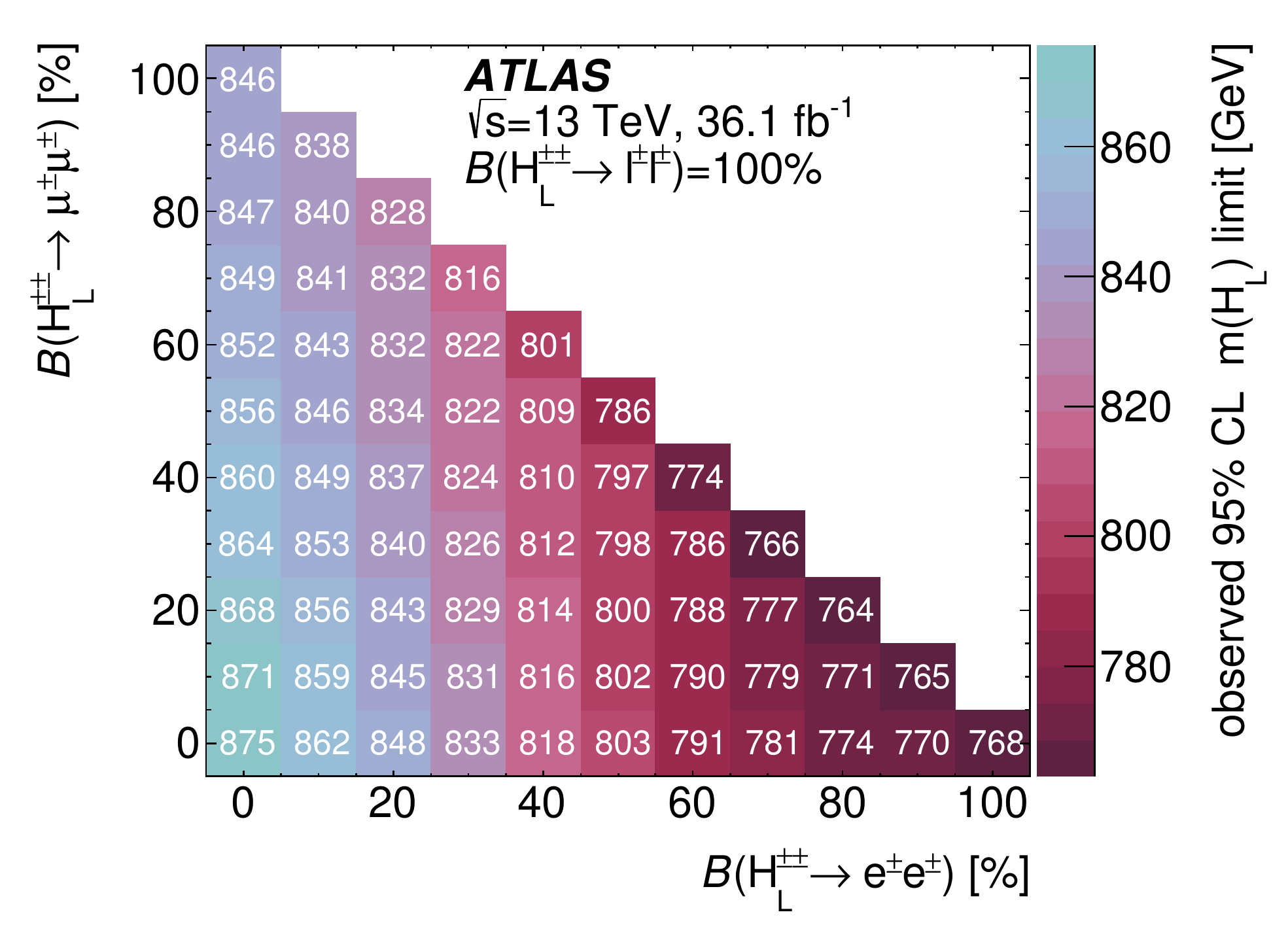}
\includegraphics[height=5.0cm]{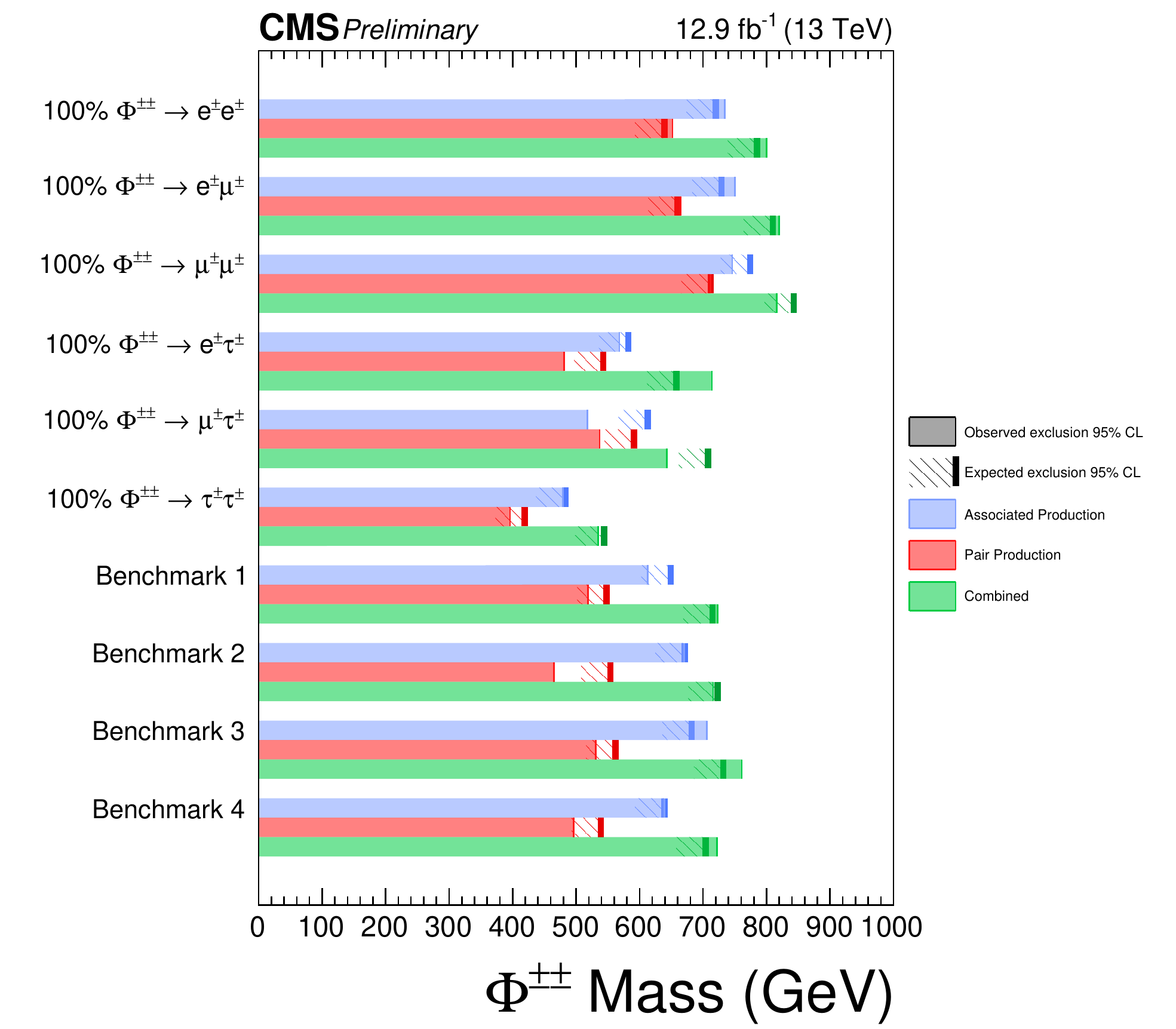}
\caption{
Limits from searches for doubly charged Higgs bosons in pair production and decays to leptons by ATLAS~\cite{atlas_hpp} (left) and CMS~\cite{cms_hpp} (middle), 
as well as in decays to W bosons by ATLAS~\cite{atlas_hpp_ww} (right).
}\label{hpp}
\end{center}
\end{figure}

For several models, the main $H^{\pm\pm}$ production mechanism at the LHC is via $\PQq\PAQq \to \PZ/\gamma^* \to \PH^{++} \PH^{--}$ or 
$\PQq\PAQq \to \PW \to \PH^{++} \PH^{-}$. Searches at the LHC focus at $H^{\pm\pm}$ decays to leptons~\cite{atlas_hpp,cms_hpp} or W bosons~\cite{atlas_hpp_ww}. 
The $H^{\pm\pm}$ coupling to leptons is typically not proportional to the lepton mass and the focus is thus on light leptons which can be handled 
more easily experimentally. The result of some of the LHC searches are illustrated in Fig.~\ref{hpp}. Depending on models and assumed branching ratios, 
typically $H^{\pm\pm}$ below 500 GeV to 900 GeV are excluded.


\section{Additional neutral Higgs bosons}
\begin{figure}[htb]
\begin{center}
\includegraphics[width=.32\textwidth]{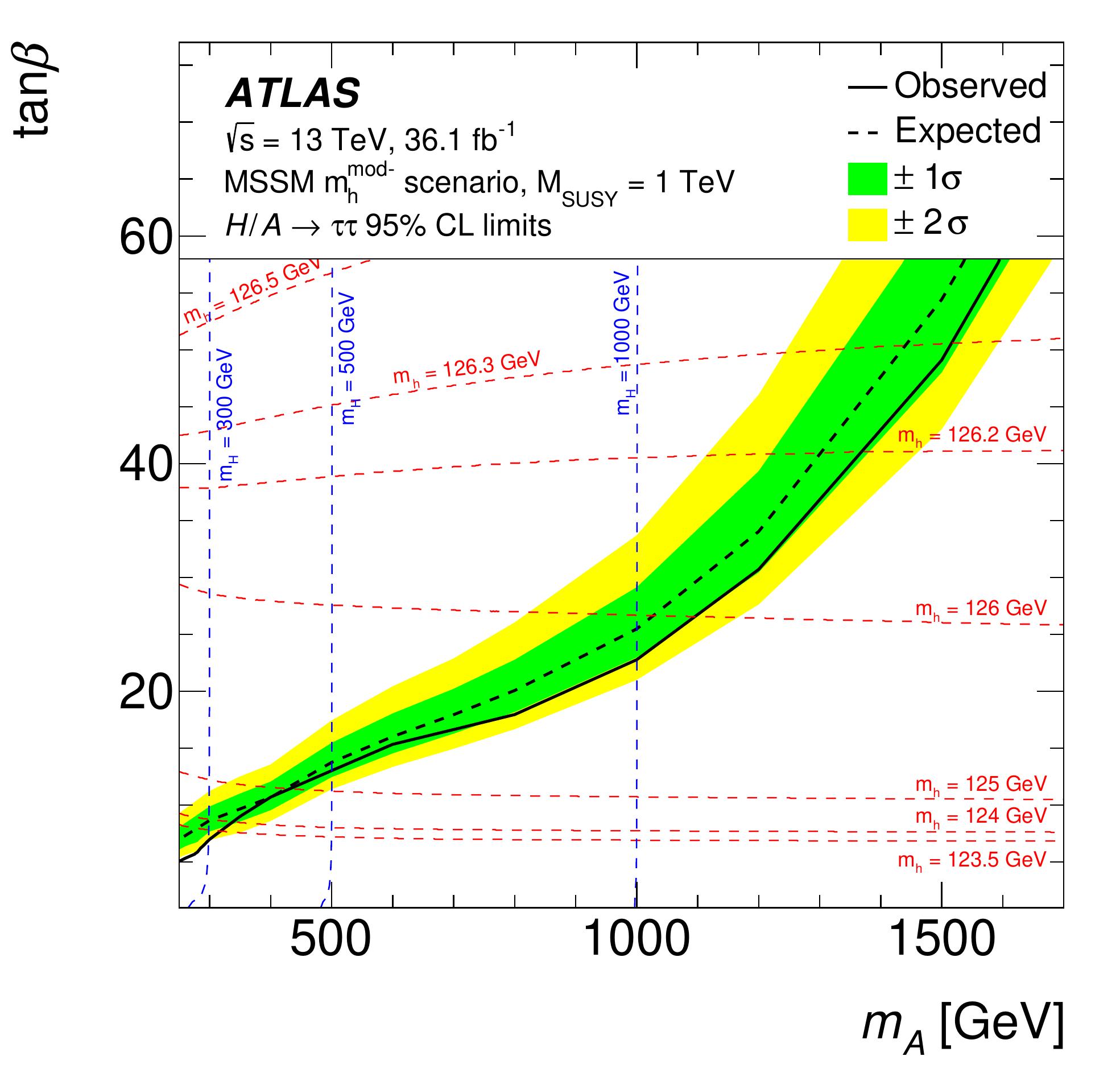}
\includegraphics[width=.32\textwidth]{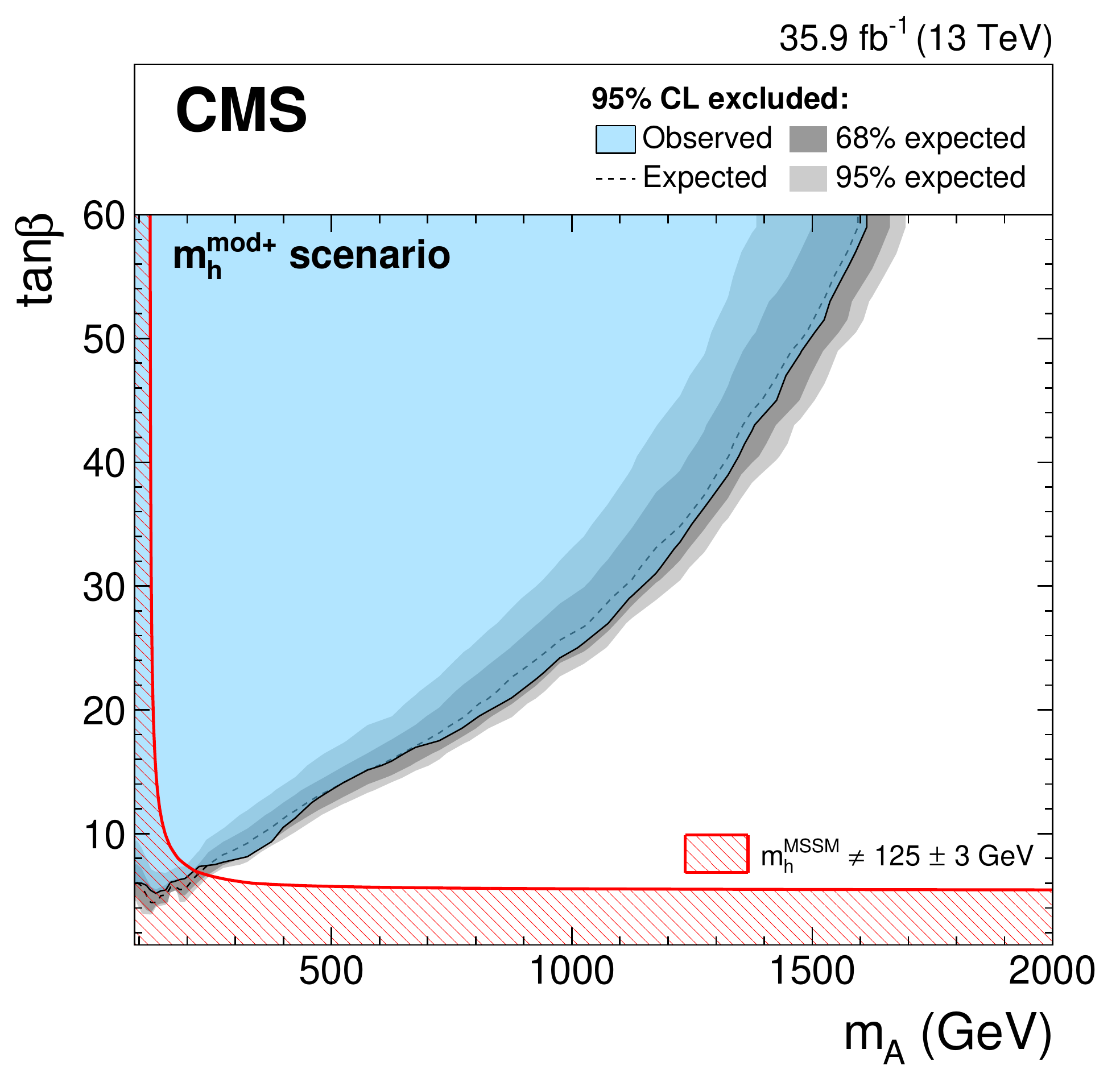}
\includegraphics[width=.32\textwidth]{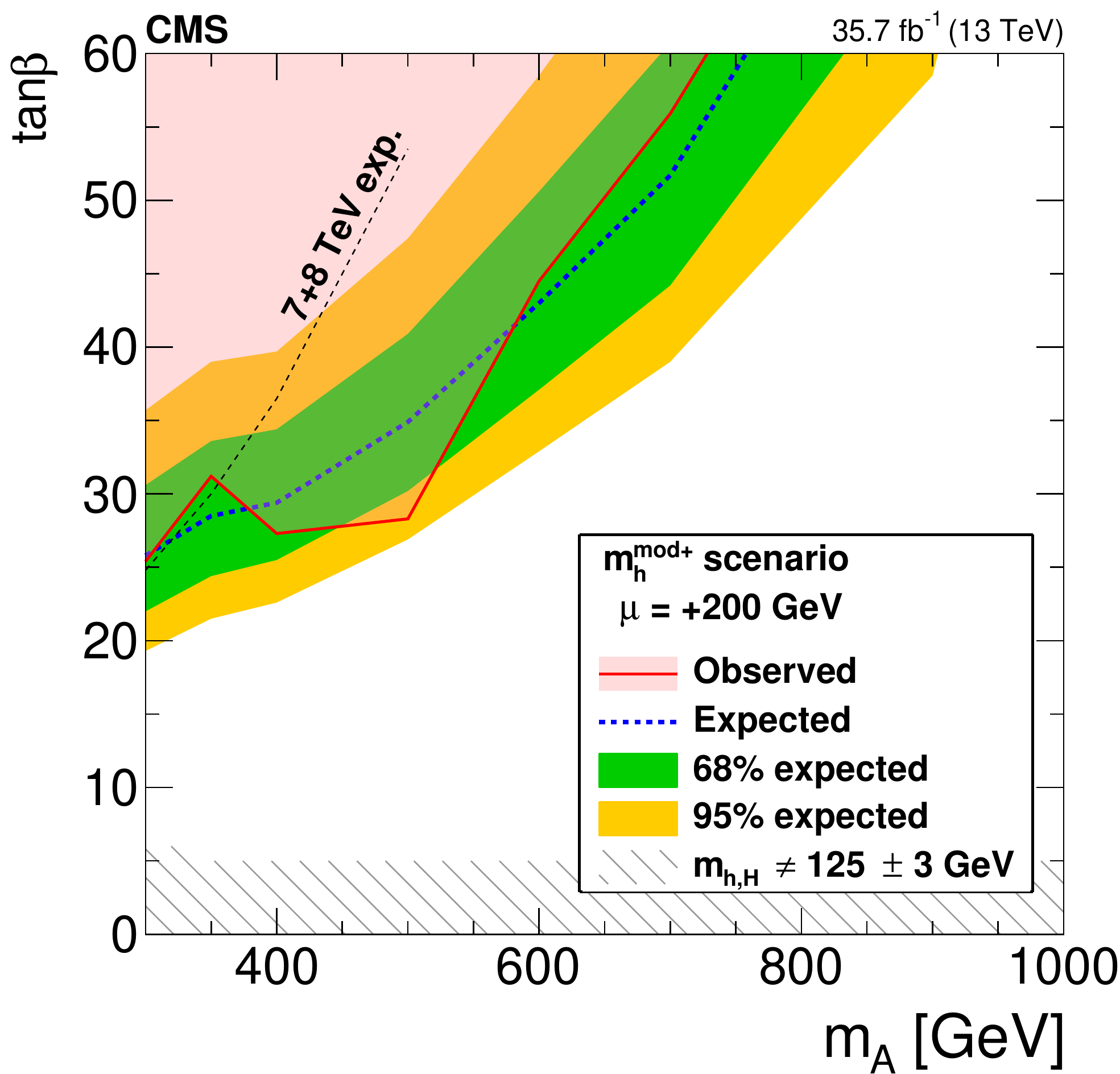}
\caption{
MSSM interpretation of the limits from searches for additional neutral Higgs bosons in $\tau\tau$ decays by ATLAS \cite{atlas_a_tau} (left) and
CMS~\cite{cms_a_tau} (middle), as well as in $\PQb\PAQb$ decays by CMS~\cite{cms_a_bb} (right).
}\label{a1}
\end{center}
\end{figure}

Additional neutral Higgs bosons are the ingredient of virtually every model with an extended Higgs sector. The main motivation for 
searches for these particles is traditionally the MSSM; however, the portfolio has been significantly extended since the LHC start. 
In addition, the MSSM searches can be reinterpreted in the context of other models. 
The main LHC production mode for neutral MSSM Higgs bosons are associated production with b quarks (intermediate and high $\tan \beta$) 
as well as gluon fusion (low $\tan \beta$). For all but very low $\tan \beta$ values, decays to $\tau$ leptons are the most sensitive 
experimental probe~\cite{atlas_a_tau,cms_a_tau} with decays to bottom quarks adding additional sensitivity~\cite{cms_a_bb}, as 
shown in Fig.~\ref{a1}. Since the last workshop, the excluded region has been significantly extended, reaching up to 1.6 TeV at high $\tan \beta$. 
At this workshop, adding the investigation of neutral Higgs boson decays to neutralinos or charginos to the portfolio~\cite{bahl} 
and investigating the impact of CP-violating effects~\cite{cp} has been proposed.

\begin{figure}[htb]
\begin{center}
\includegraphics[width=.32\textwidth]{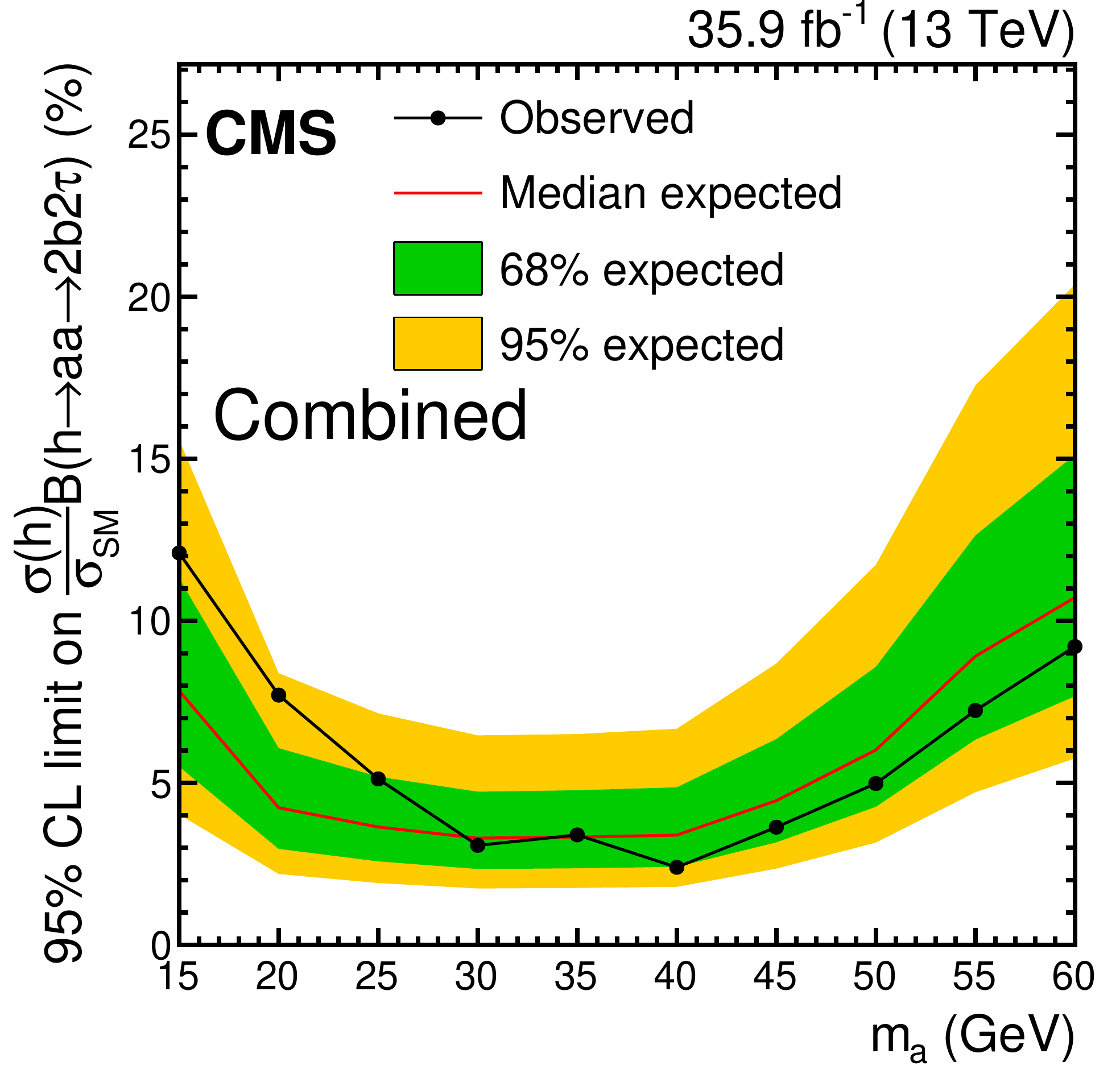}
\includegraphics[width=.32\textwidth]{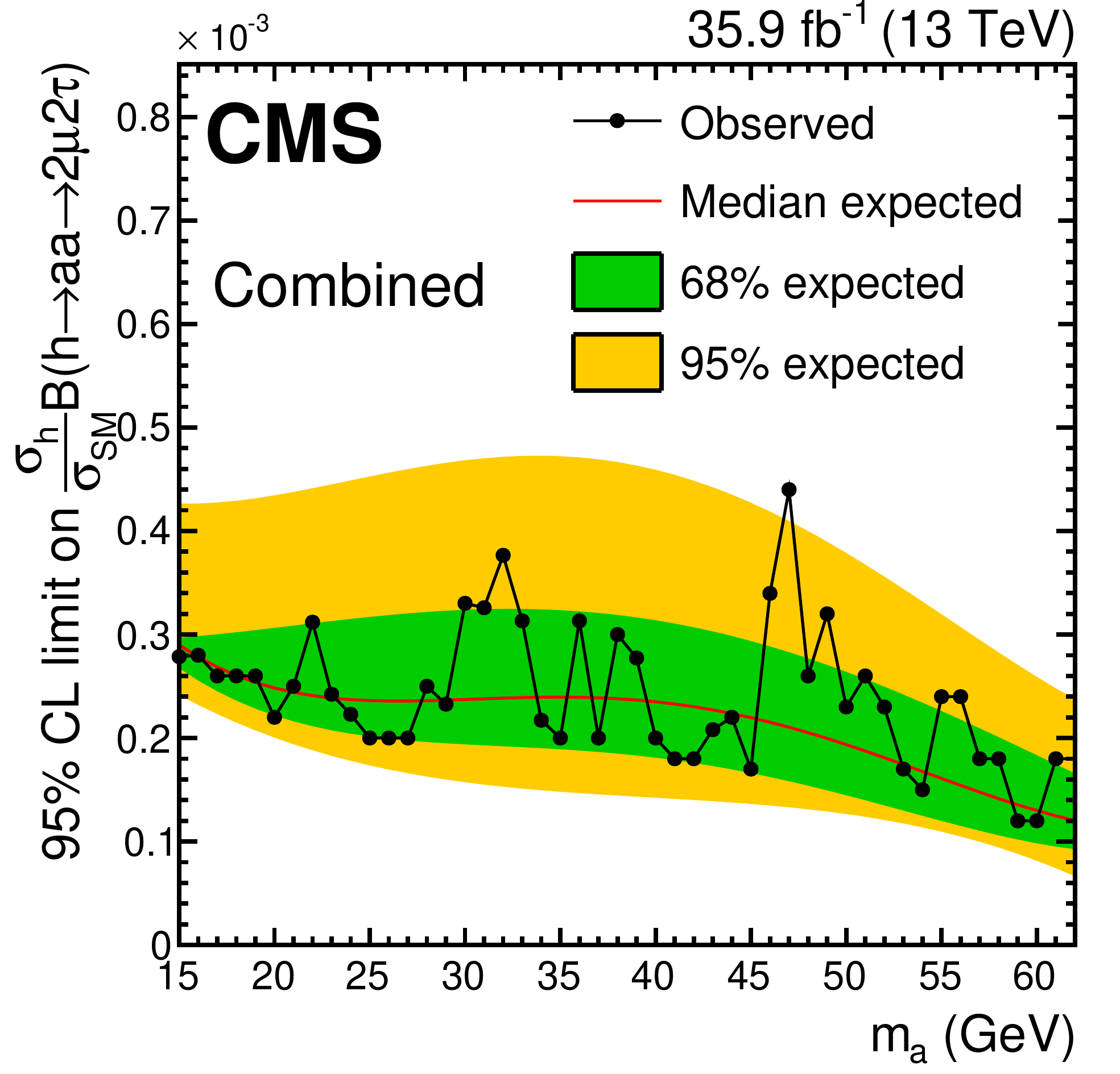}
\caption{
95\% CL cross section limits on the cross section times branching ratio 
of $\Ph \to \Pa\Pa$ in the $2\PQb 2\tau$~\cite{cms_aa_2b2t} and $2\mu 2\tau$~\cite{cms_aa_2m2t} final states.
}\label{aa}
\end{center}
\end{figure}

In models with additional Higgs bosons with masses below about 62.5 GeV, the 125-GeV Higgs boson can decay to these Higgs bosons, $\Ph \to \Pa\Pa$. 
Searches for these light Higgs bosons are often motivated by the NMSSM but the results can be interpreted in a generic way. These bosons 
then dominantly decay to the heaviest particles kinematically allowed, leading to a variety of search modes~\cite{cms_aa_2b2t,cms_aa_2m2t}. 
The results of two such searches are shown in Fig.~\ref{aa}. 

\begin{figure}[htb]
\begin{center}
\includegraphics[height=4.5cm]{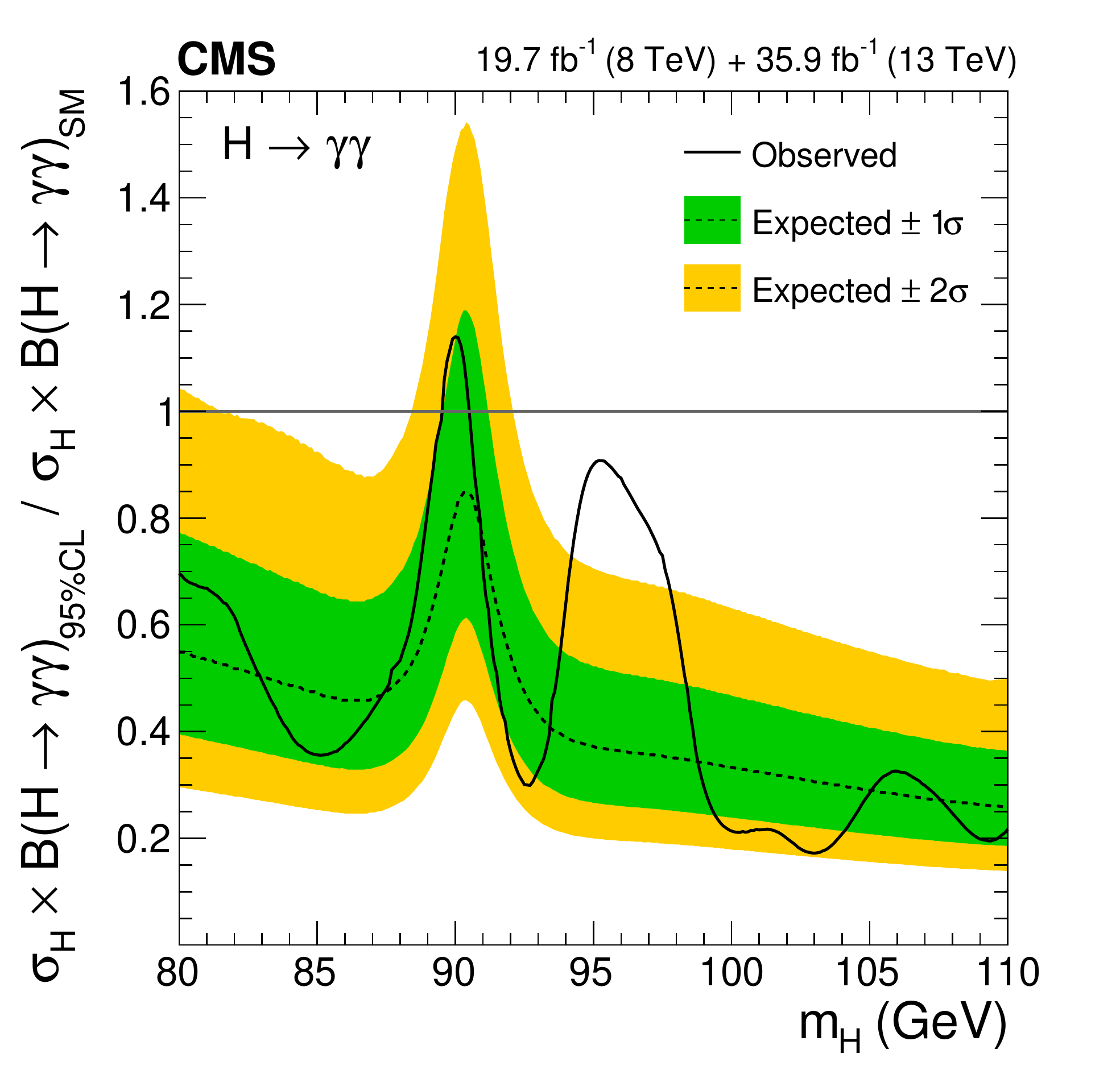}
\includegraphics[height=4.5cm]{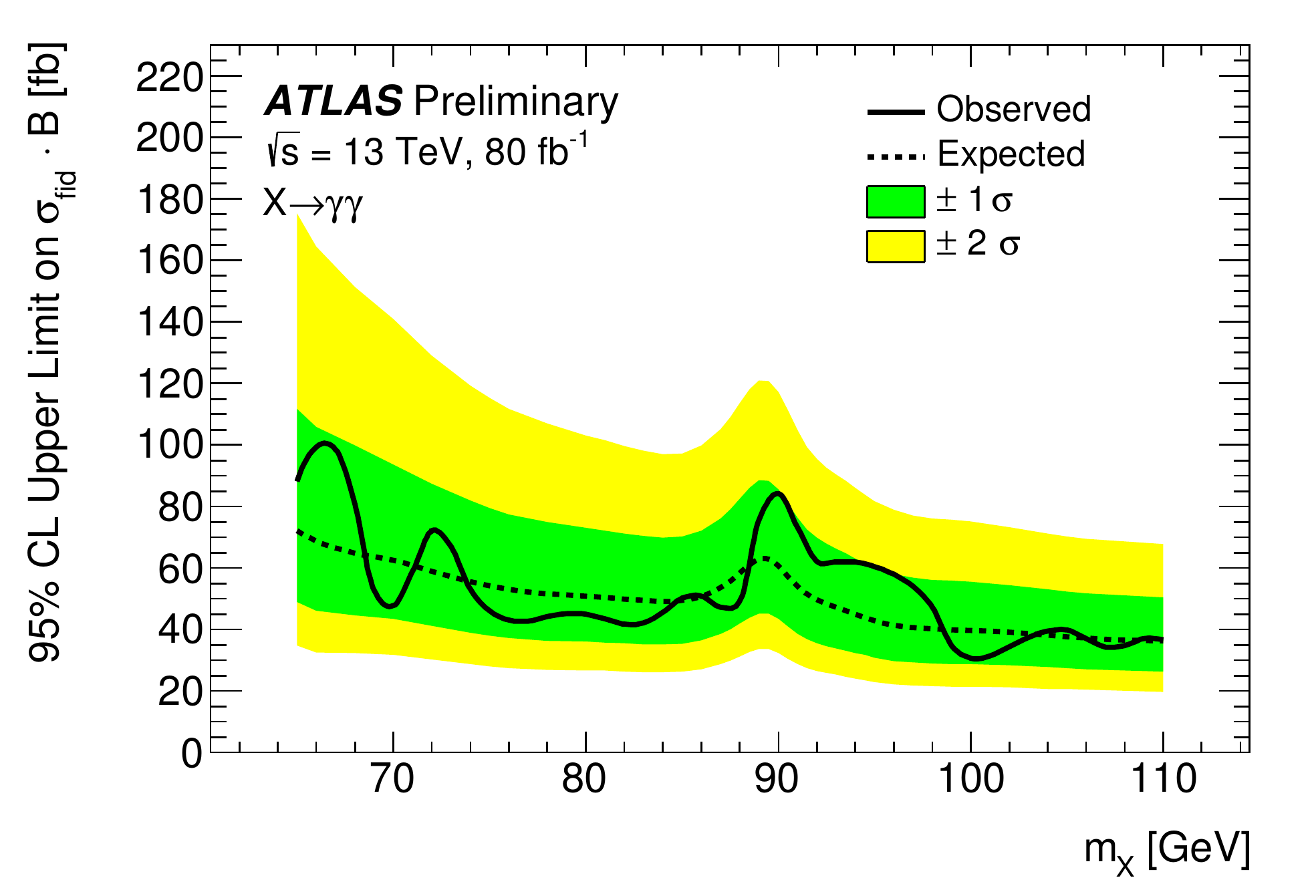}
\caption{
95\% CL cross section limits on light Higgs bosons in $\gamma\gamma$ decays by CMS~\cite{cms_lowmass} (left) and ATLAS~\cite{atlas_lowmass} (right). 
}\label{lowmass}
\end{center}
\end{figure}

\begin{figure}[htb]
\begin{center}
\includegraphics[height=4.7cm]{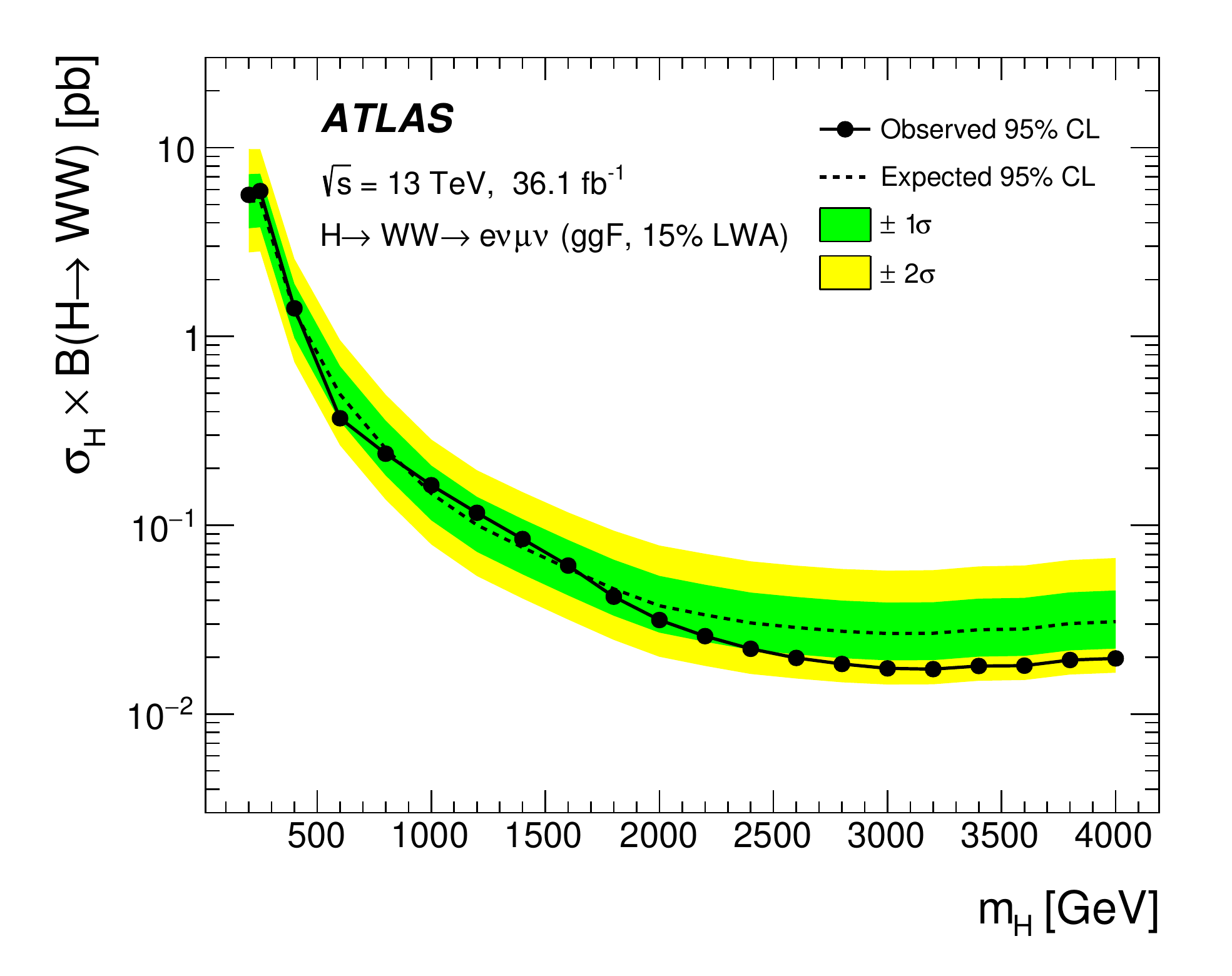}
\includegraphics[height=4.7cm]{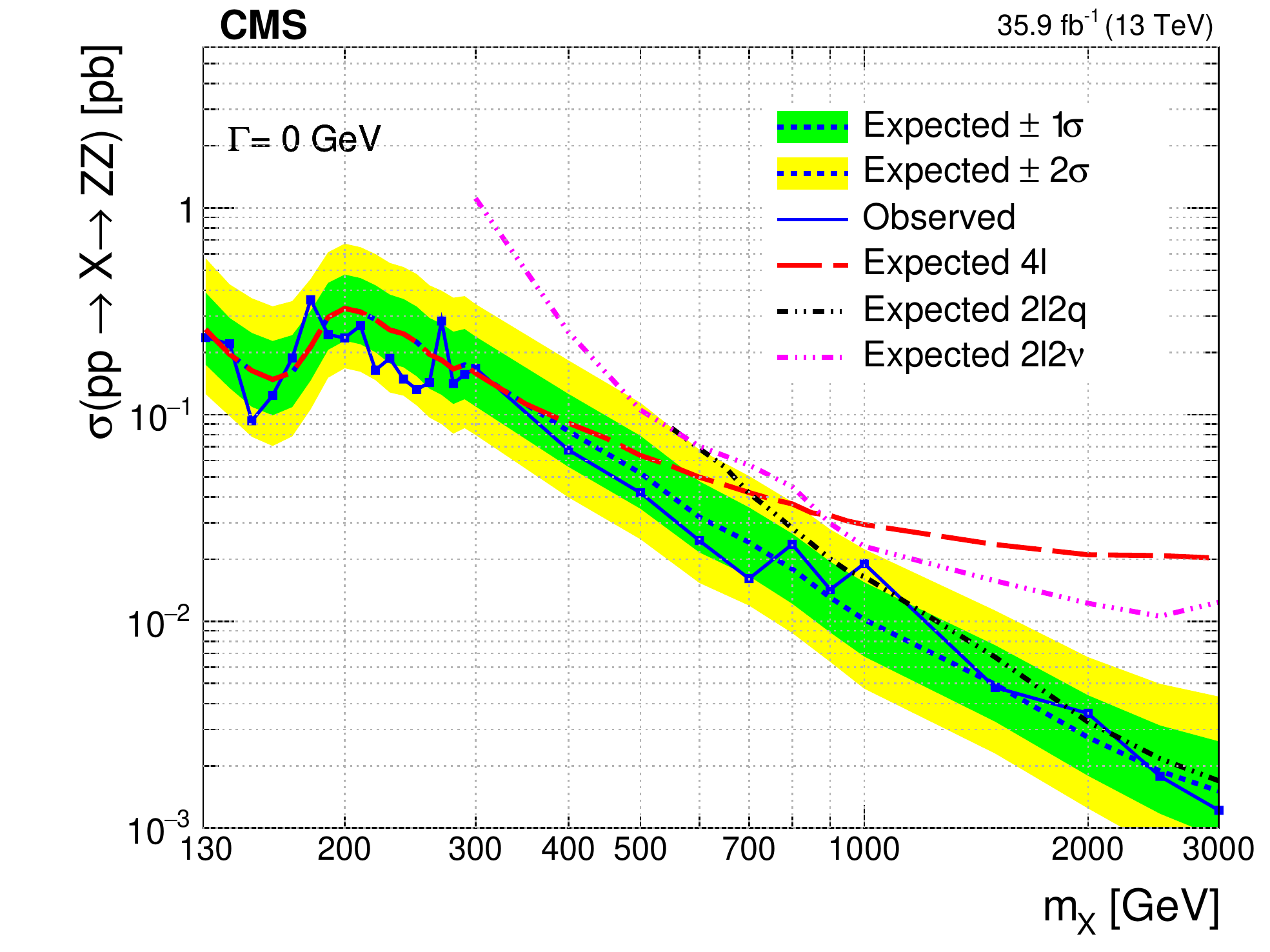}
\caption{
95\% CL cross section limits on heavy Higgs bosons in VV decays by ATLAS~\cite{atlas_ww} (left) and CMS~ \cite{cms_zz} (right).
}\label{vv}
\end{center}
\end{figure}

Generically, light Higgs bosons are searched for in $\gamma\gamma$ decays in 
the mass range 65 GeV--110 GeV. Here, a slight excess of events is observed for $m_{\gamma\gamma} \approx 95$ by CMS~\cite{cms_lowmass} which 
is not confirmed by ATLAS~\cite{atlas_lowmass}, see Fig.~\ref{lowmass}. Similarly, generic Higgs boson searches are extended to high masses 
up to 4 TeV in the WW~\cite{atlas_ww} and ZZ~\cite{cms_zz} decay modes. At the highest masses, a cross section as low as 1 fb is excluded, see 
Fig.~\ref{vv}.


\section{Resonant decays to Higgs bosons}
\begin{figure}[htb]
\begin{center}
\includegraphics[width=.4\textwidth]{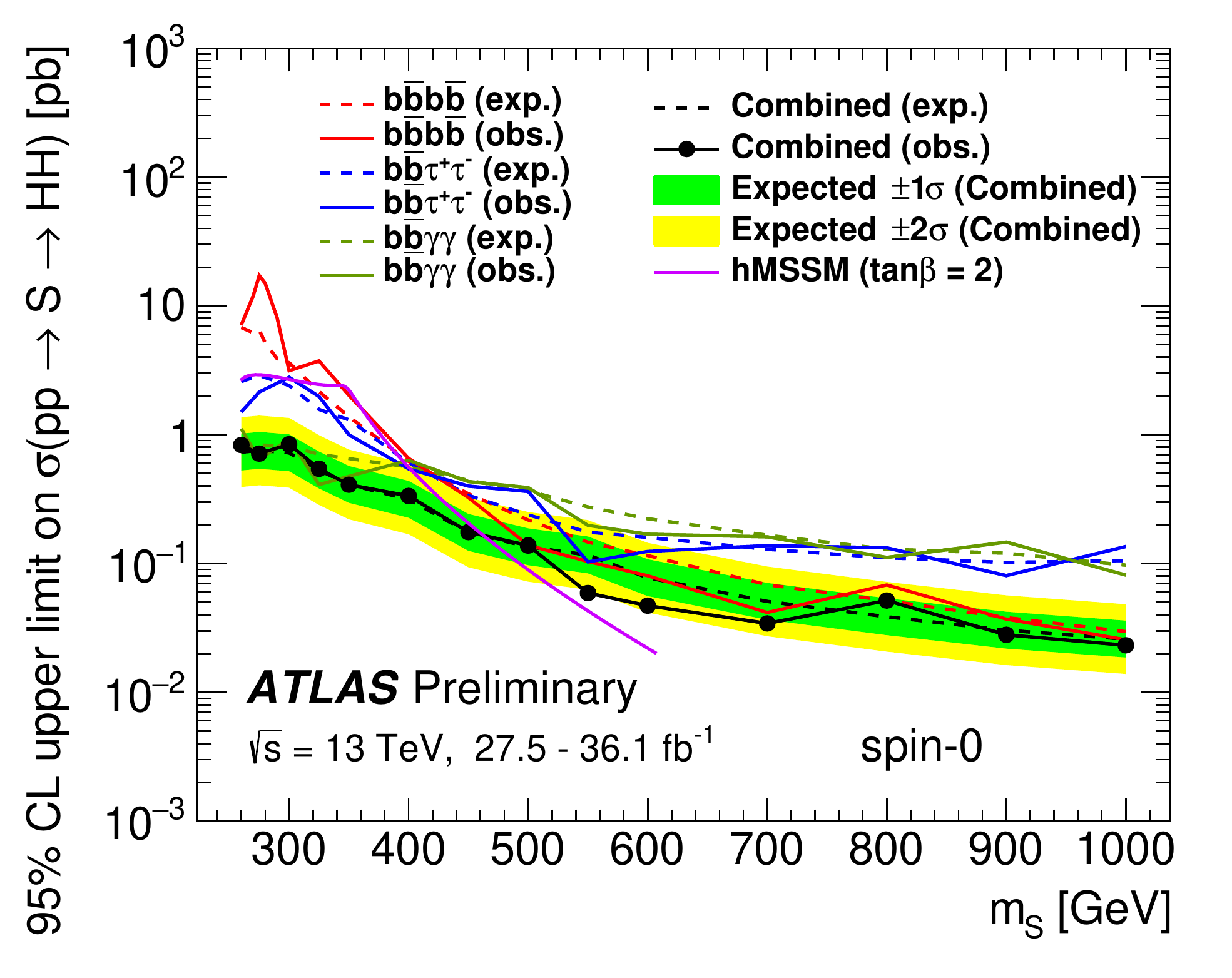}
\includegraphics[width=.5\textwidth]{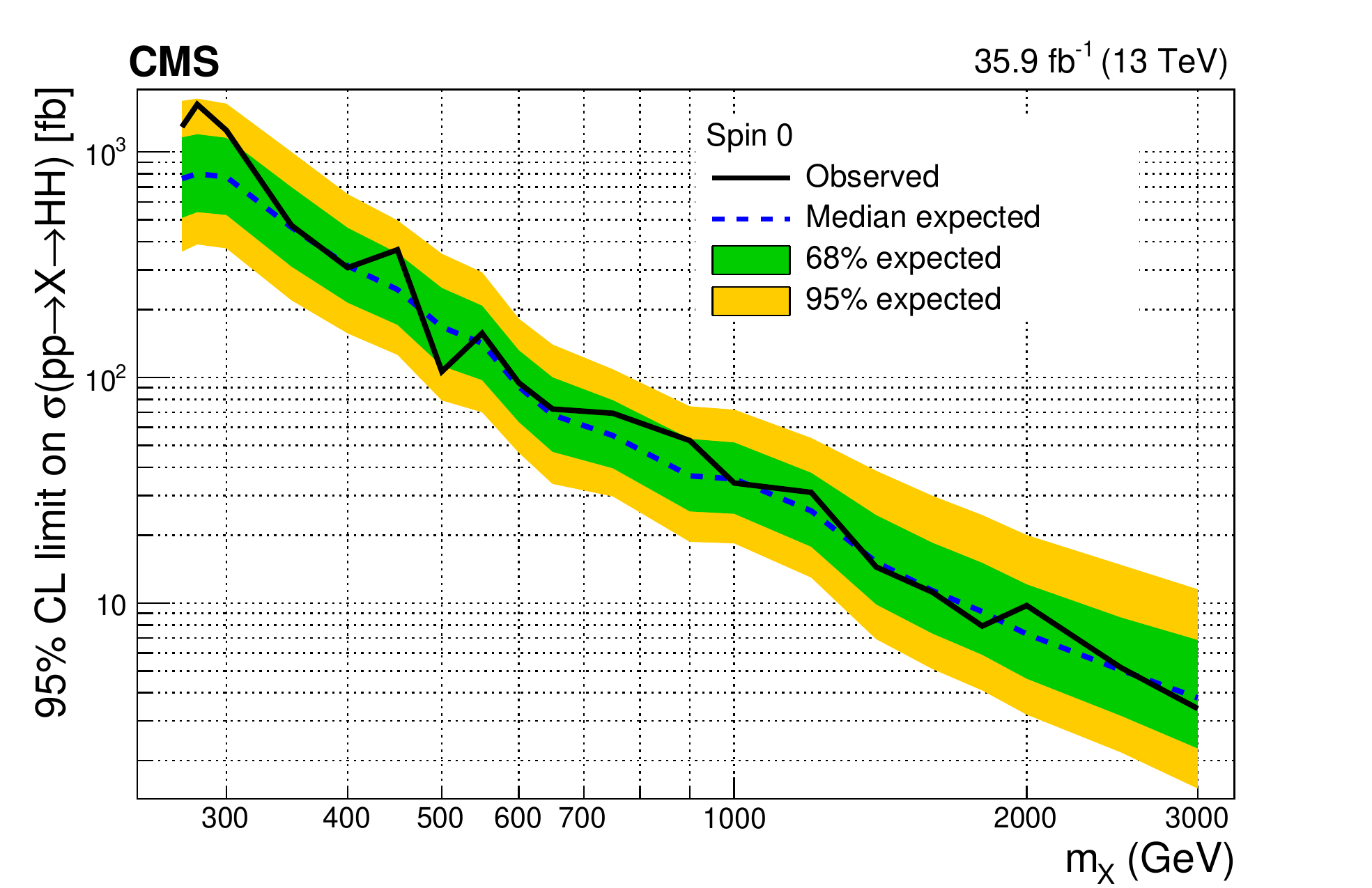}
\caption{
95\% CL cross section limits on a scalar resonance decaying to two Higgs bosons with $m_{\PH}=125$ GeV for ATLAS~\cite{atlas_hh} (left) and CMS~\cite{cms_hh} (right).
}\label{hh}
\end{center}
\end{figure}

The SM predicts the production of events with two Higgs bosons of a mass of 125 GeV with a relatively low pp cross section at $\sqrt{s}=13$ TeV 
of about 33 fb, including diagrams both with and without vertices with three Higgs bosons. This has not been observed yet at the LHC and 
the current 95\% CL limits, in multiples of the expected SM cross section, are 6.7 observed (10.4 expected) for ATLAS~\cite{atlas_hh} 
and 22.2 observed (12.8 expected) for CMS~\cite{cms_hh}. An increased rate of HH production could be due to a heavy resonance X
decaying to two Higgs bosons, i.e. $\Pp\Pp\, \rightarrow\, \mathrm{X}\, \rightarrow\, \PH \PH$. Both ATLAS and 
CMS have searched for such resonant decays but found good agreement with the SM expectation. Limits on a scalar resonance decaying to 
HH, combining several final states, are shown in Fig.~\ref{hh}, and reach from about 1 pb at $m_\mathrm{X}=300$ GeV down to a few fb at $m_\mathrm{X}=3000$ GeV.


\section{BSM search summary}
\begin{figure}[htb]
\begin{center}
\includegraphics[height=5.9cm]{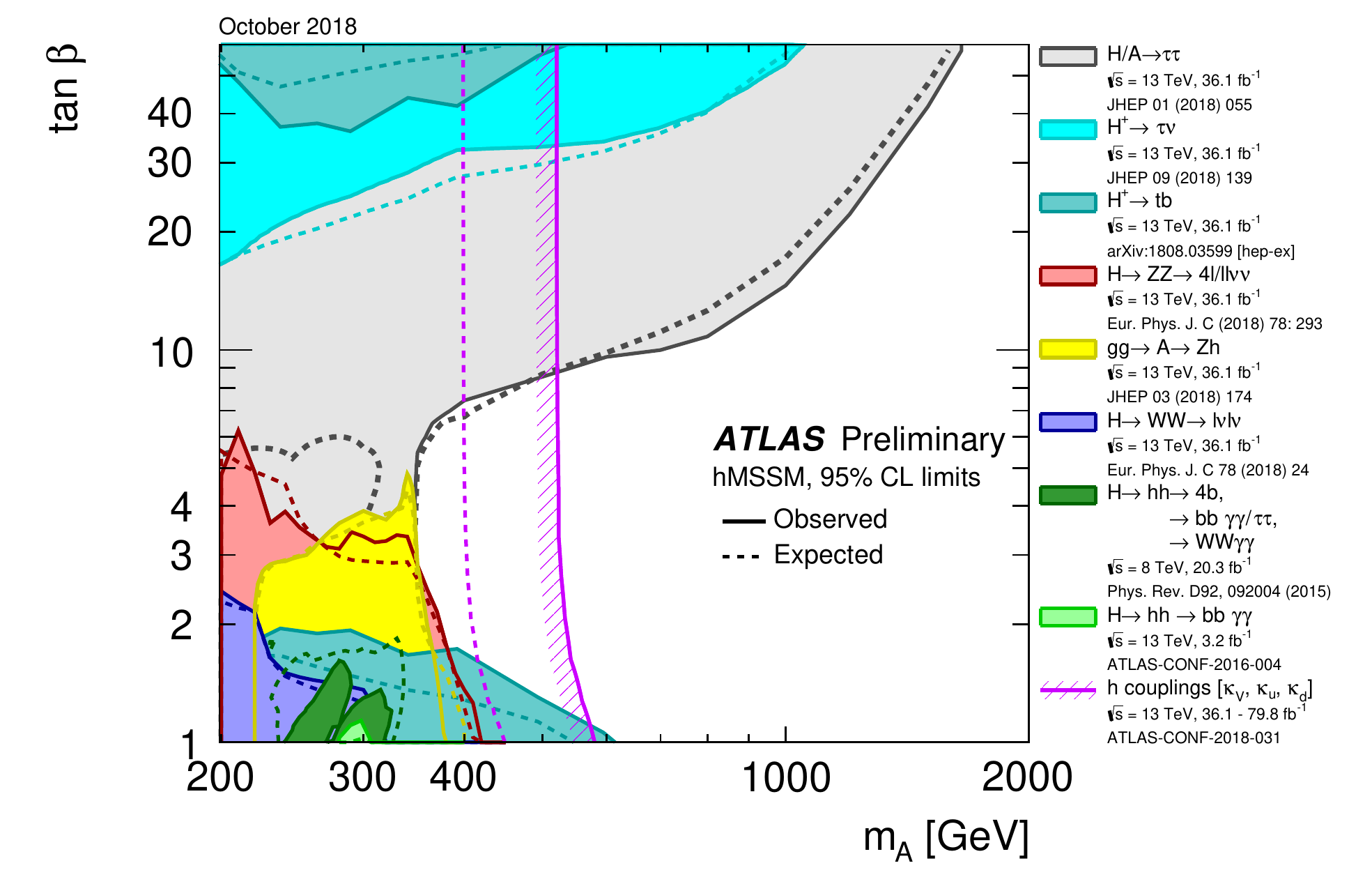}
\includegraphics[height=5.8cm]{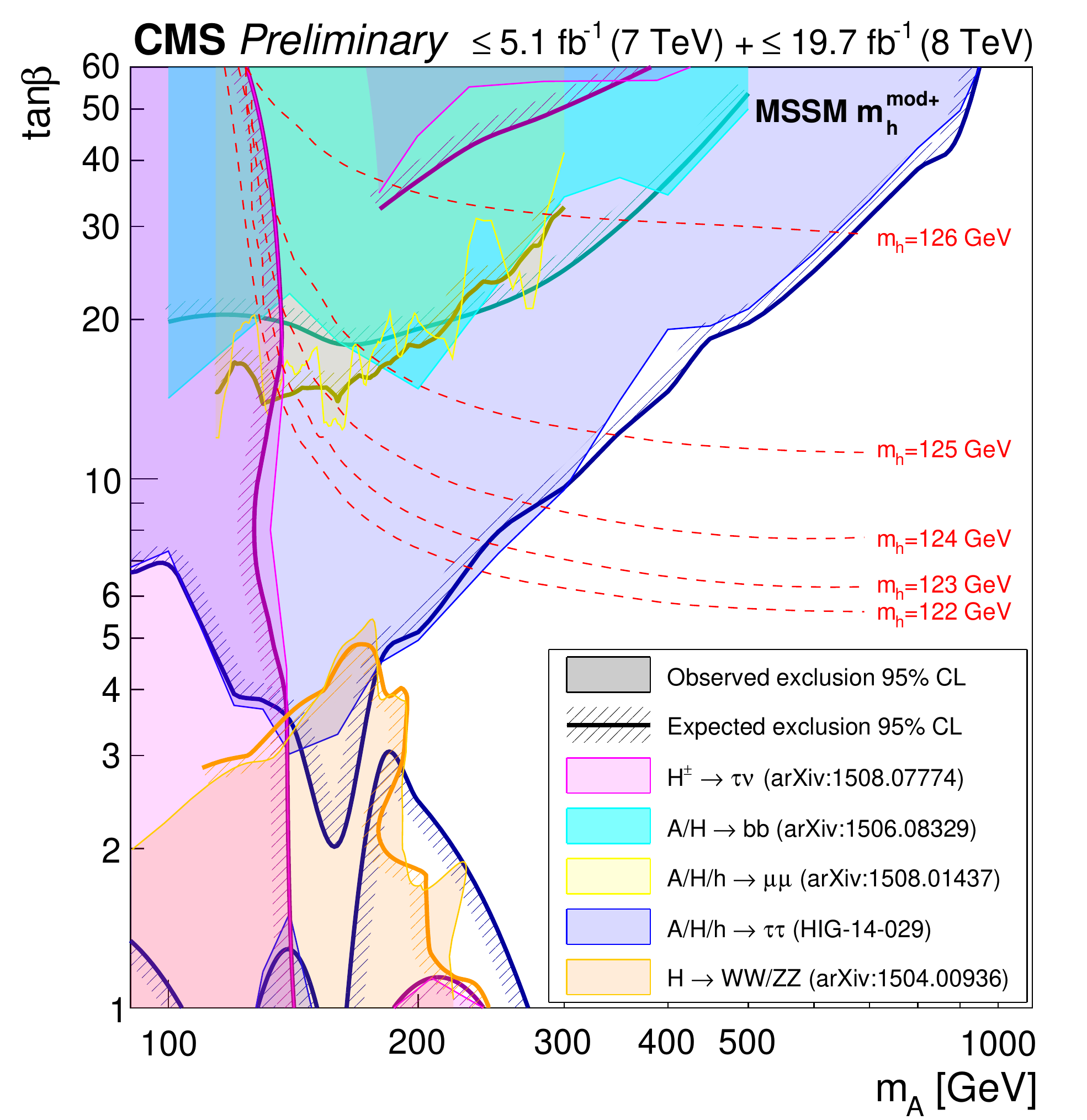}
\caption{
Summary plots of BSM Higgs boson searches for the hMSSM scenario by ATLAS~\cite{atlas_sum} and for the $\mhmodp$ scenario by CMS~\cite{cms_sum}.
}\label{sum}
\end{center}
\end{figure}

For given benchmark scenarios, limits using different final states can be compared which is particularly useful if they have their main sensitivity in 
different regions of the parameter space. Examples for the MSSM scenarios hMSSM (ATLAS)~\cite{atlas_sum} and $\mhmodp$ (CMS)~\cite{cms_sum} are given 
in Fig.~\ref{sum}. The most stringent limits are obtained for the hMSSM where an additional neutral Higgs boson with $m_{\PH}<500$ GeV is excluded 
(e.g. by using, for each point in parameter space, the observed limit corresponding to the most sensitive analysis there). While analyses of the $\tau\tau$ 
final state cover the excluded parameter space at intermediate and high $\tan \beta$, decays involving bosons or up-type fermions are needed to close 
the gap at low $\tan \beta$.

\section{Prospects at future colliders}
\begin{figure}[htb]
\begin{center}
\includegraphics[height=7cm]{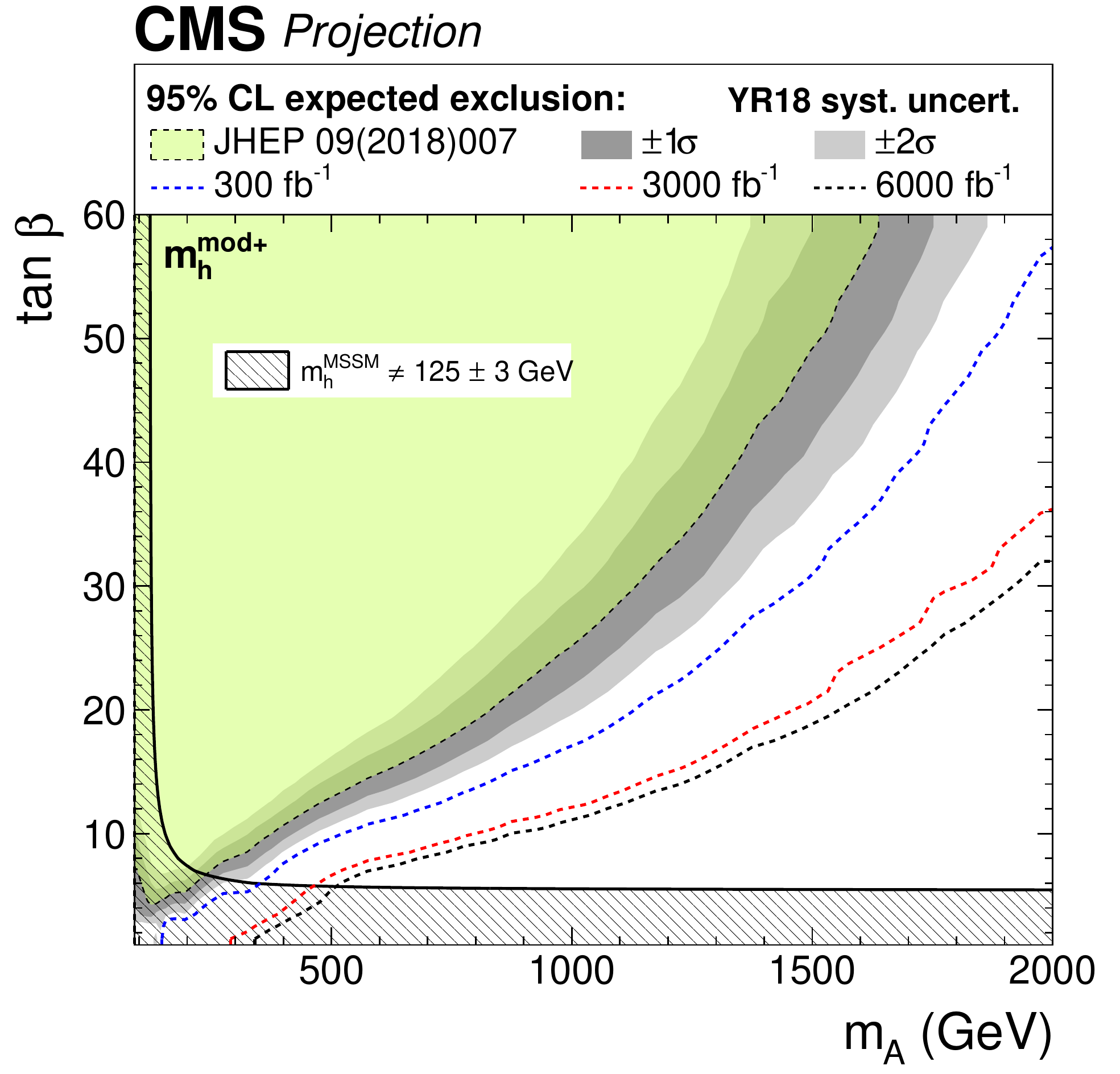}
\includegraphics[height=7cm]{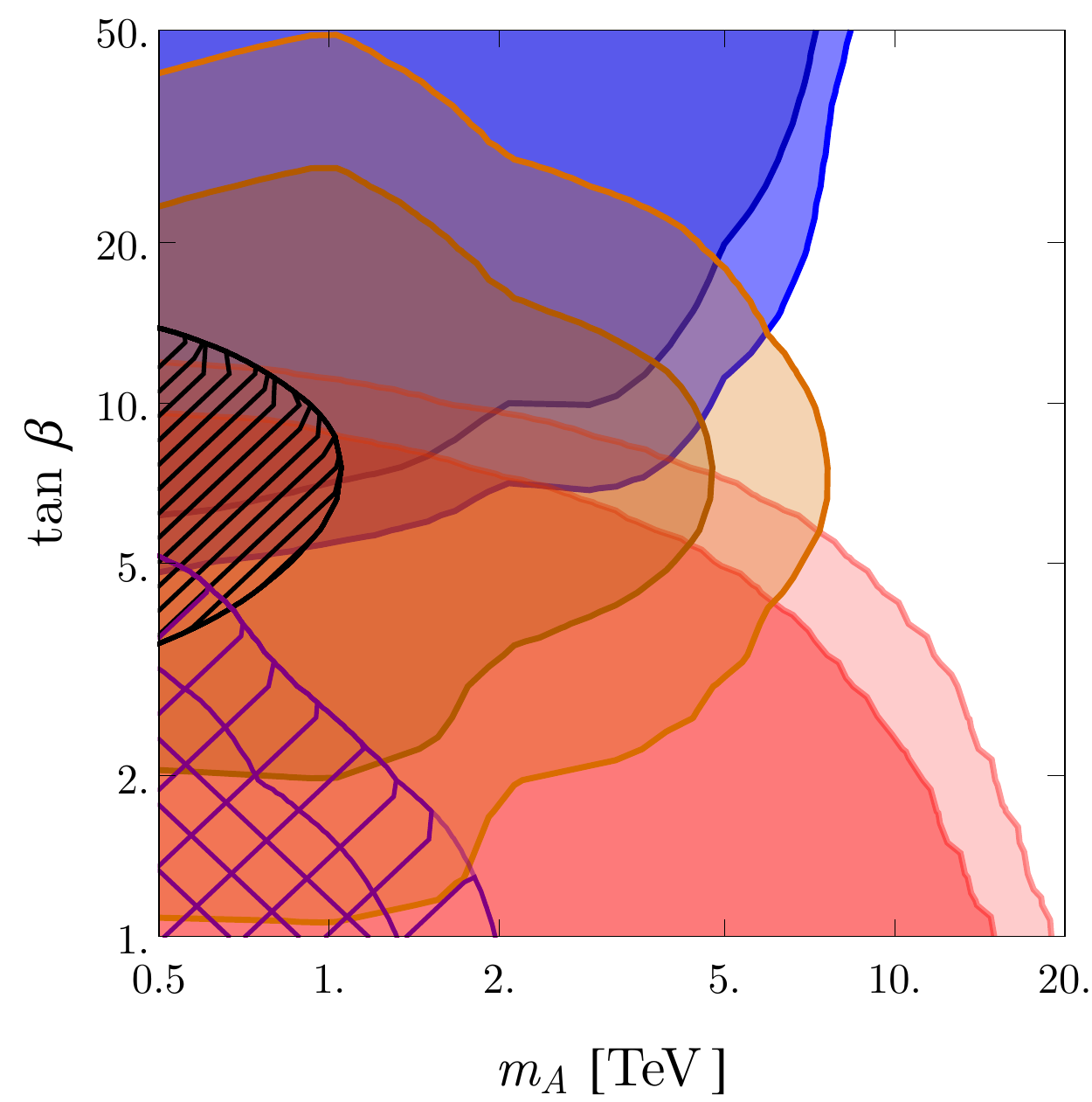}
\caption{
Exclusion sensitivity for neutral heavy Higgs bosons. Left: High-Luminosity LHC~\cite{cms_proj_a} reach for $\PSA \to \tau\tau$.
Right:
The low $\tan\beta$ region (red) is covered by $\PQt \PAQt \PH$, $\PH \to \PQt \PAQt$.
The intermediate $\tan\beta$ region (orange) is covered by $\PQb$-associated production with $\PH \to \PQt \PAQt$ decays. 
The large $\tan\beta$ region (blue) is covered by $\PQb$-associated production with $\tau\tau$ decays.
The smaller bound (lines) corresponds to 0.3 and 3 ab$^{-1}$, the larger bound (filled regions) to 3 and 30 ab$^{-1}$ at the LHC and a future pp-collider, respectively~\cite{fcc_proj_a}.
}\label{proj}
\end{center}
\end{figure}

Sensitivity projections for neutral MSSM Higgs bosons are shown in Fig.~\ref{proj} for the LHC and a future pp-collider. While the LHC will be able to 
probe a region up to 2 TeV at intermediate and high $\tan \beta$ using $\tau\tau$ decays of the Higgs boson, a future collider will be able to probe the region up 
to about 5 TeV for all $\tan \beta$, combining the information from different final states.

\section{Summary}
While so far none of the searches for additional Higgs bosons revealed a significant deviation from the SM expectation, the remaining parameter space has 
been strongly limited in the last two years with an increased amount of analyzed LHC data at $\sqrt{s}=13$ TeV, and the increasing number of models and 
final states investigated. The LHC and future colliders will further push these boundaries in the coming decades, having the sensitivity to discover 
additional Higgs bosons.


\end{document}